\documentclass[apex]{jjap3}
\usepackage{bm}

\begin{document}

\title{Degree of Order Dependence on Magnetocrystalline Anisotropy\\in Bct FeCo Alloys\footnote{Submitted 19 September 2012; Accepted 15 October 2012 in Applied Physics Express.}}
\author{Yohei Kota\thanks{E-mail address: kota@solid.apph.tohoku.ac.jp} and Akimasa Sakuma}
\inst{Department of Applied Physics, Tohoku University, Sendai 980-8579, Japan}
\abst{We investigate the magnetocrystalline anisotropy (MCA) energy of tetragonal distorted FeCo alloys depending on the degree of order by first-principles electronic structure calculation combined with the coherent potential approximation. The obtained results indicate that the MCA energy of FeCo alloys strongly depends on the degree of order under optimal conditions, where the axial ratio of the bct structure is 1.25 and the composition is Fe$_{0.5}$Co$_{0.5}$. We find that the modification of the electronic structure resulting from electron scattering by chemical disorder has a considerable influence on the MCA under these conditions.
}

\makeatletter
\makeatother

\maketitle

In recent elemental strategies, it has been essential for hard magnetic materials applied to magnetic devices and permanent magnets to be free from rare-metal and rare-earth elements. Given this situation, Burkert {\it et al.} proposed that the FeCo alloy can be expected to exhibit a giant magnetocrystalline anisotropy (MCA) energy under certain conditions by the first-principles calculation using the virtual crystal approximation (VCA) to treat the binary alloy system \cite{burkert2004}. In their paper, tetragonal distorted Fe$_{1-x}$Co$_{x}$ alloys exhibit a large uniaxial magnetic anisotropy constant $K_\mathrm{u}$, as high as about 700 -- 800 $\mu \text{eV/atom}$, which is sufficiently comparable to the $K_\mathrm{u}$ value of a series of  $L1_0$-type alloys such as FePt, CoPt, and FePd,\cite{klemmer1995} under the conditions that the axial ratio between the $a$ and $c$ axes, $c/a,$ is 1.20--1.25 and that the composition $x$ is 0.5--0.6. This result has stimulated considerable expectation for their use as novel materials with high coercivity; thus, many experimental trials have been performed to realize such a property \cite{andersson2006,winkelmann2006,warnicke2007,luo2007,yildiz2009}. Under the optimal conditions, the magnetic easy axis coincides with the $c$-direction  in agreement with the theoretical prediction; however, satisfactory results have not yet been obtained at a quantitative level.

Recently, Neise {\it et al.} \cite{neise2011} have performed a first-principles investigation on the MCA energy of FeCo alloys with a chemical disordered structure by employing both the VCA and the supercell method. When using the VCA, they obtained  results  similar to those of  Burkert {\it et al.}\cite{burkert2004}; however, for the supercell structure reflecting chemical disorder, they obtained  values of the MCA energy smaller by factors of 1.5--3.0. Therefore, we can consider that one of the reasons for the disagreement with the experiments may be the chemical disorder. Since the VCA is generally the lowest order approximation in terms of a random configuration of atoms, it may be inappropriate to calculate the MCA energy in FeCo owing to the insufficient treatment of the chemical disorder \cite{hautier2012}. However, the physical explanation for this discrepancy and the influence of the chemical disorder on the MCA in FeCo are not yet understood. In this work, we investigate the MCA in disordered FeCo alloys by means of the coherent potential approximation (CPA) \cite{velicky1968,elliott1974}, which provides us with a physical picture of the effects of chemical disorder in a transparent manner. Employing the CPA has the advantage of reflecting electron scattering by the chemical disorder, which has not been considered in the VCA and the supercell approach. In addition, it  allows us to continuously change the order parameter $S$ of the crystal \cite{staunton2004,kota2012a,kota2012b}, a parameter that is measurable in experiments.

Electronic structure calculations are performed by using the tight-binding linear muffin-tin orbital (TB-LMTO) method under the local spin-density functional approximation \cite{skriver1984,turek1997,turek2008}. Chemical disorder is taken into account by means of both the CPA and the lowest order VCA to compare the obtained results. In the former approximation, the averaged potential function $\langle P \rangle$, which is used in the TB-LMTO method and reflects the random distribution of Fe and Co atoms, is determined via the iterative CPA condition for single-site $t$-matrices. In the latter, $\langle P \rangle$ is determined only by the simple weighted average of the potential functions of  both atoms in accordance with the composition ratio \cite{turek1997}, so there is no electron scattering. In both methods, we take advantage of the Green's function technique, which needs an infinitesimal imaginary value reflecting causality; we set this value to 2.0 mRy in this work. The MCA energies are calculated based on the force theorem with inclusion of the spin-orbit interaction after a self-consistent calculation. The lattice constants are determined by the volumes of the unit cell of the experimental data of the bulk Fe-Co alloys \cite{bozorth1951} and the optional axial ratio $c/a$. The number of $k$-points used for MCA energy calculation is about $50^3$ in the full Brillouin zone.

\begin{figure}
\begin{center}
\includegraphics[clip,width=6cm]{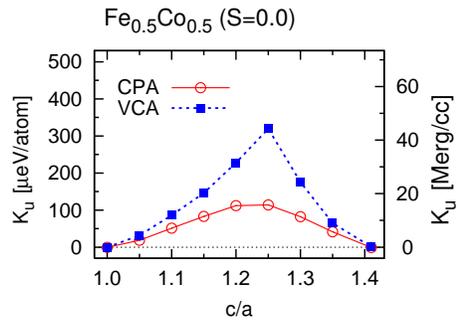}
\end{center}
\caption{Uniaxial anisotropy constant $K_\mathrm{u}$ of FeCo for $S=0.0$ as a function of the axial ratio $c/a$. The results obtained from the CPA and VCA are denoted by the open circles and filled squares, respectively.}
\label{fig1}
\end{figure}

First, we examine the $K_\mathrm{u}$ value of completely disordered Fe$_{0.5}$Co$_{0.5}$ $(S=0.0)$ as a function of the axial ratio $c/a$ in Fig. \ref{fig1}. In  both the CPA and the VCA, we can measure the peak of $K_\mathrm{u}$ around $c/a=1.25$. We note that $K_\mathrm{u}$ is zero for $c/a=1.0 \ \text{and} \ \sqrt{2}$ owing to the cubic symmetry. This behavior is qualitatively consistent with the results of Burkert {\it et al.} \cite{burkert2004}, although the value of $K_\mathrm{u}$ calculated using the VCA is almost half at the peak position. This discrepancy may be attributed to the difference in the band calculation method; Burkert {\it et al.} \cite{burkert2004} used the full-potential LMTO method, which can provide more accurate results than the present method if one calculates the electronic structure of a perfectly ordered crystal, {\it i.e.}, $S=1.0$. Here, it should be emphasized that the value of $K_\mathrm{u}$ calculated using the CPA is further reduced compared with that calculated using the VCA, especially around $c/a=1.25$. This behavior is consistent with the results obtained by Neise {\it et al.} using the supercell method \cite{neise2011}.

\begin{figure}
\begin{center}
\includegraphics[clip,width=6cm]{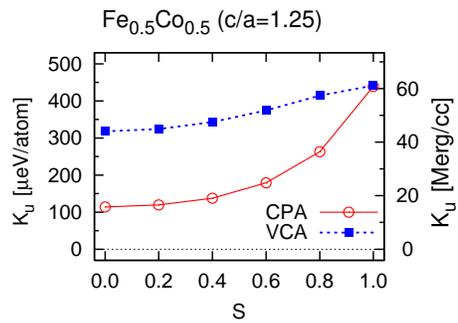}
\end{center}
\caption{Uniaxial anisotropy constant $K_\mathrm{u}$ as a function of degree of order $S$ of FeCo with  $c/a=1.25$. The open circles and filled squares represent the results from the CPA and VCA, respectively.}
\label{fig2}
\end{figure}

Next, let us look at the degree of order dependence of $K_\mathrm{u}$ for $c/a=1.25$ in Fe$_{0.5}$Co$_{0.5}$ in Fig. \ref{fig2}. Here, we note that there are two equivalent sites in the primitive cell of the tetragonal lattice. We set the composition at the body center site and the corner one in the primitive cell to Fe$_{1-\eta}$Co$_{\eta}$ and Fe$_{\eta}$Co$_{1-\eta}$ $(0<\eta<0.5)$, respectively;  the long-range order parameter of the crystal is defined by $S=1-2\eta$ ($0.0<S<1.0$) \cite{kota2012b}. For $S=1.0$, the lattice is a perfectly ordered structure; therefore, the results obtained  using the VCA and CPA are invariant. In Fig. \ref{fig2}, there is a large difference between these results  for $S<0.8$; therefore, we can consider that the values obtained using the lowest order VCA possibly overestimate the uniaxial anisotropy constant of FeCo compared with the values obtained using the CPA, which is a more proper way of dealing with a disordered alloy system \cite{elliott1974}. Therefore, this result reveals that the giant MCA expected from FeCo is not present even if there is a slight disorder in the crystal.

\begin{figure}
\begin{center}
\includegraphics[clip,width=4.5cm]{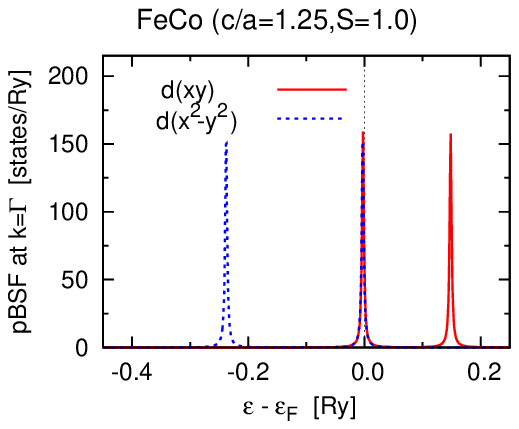} \\ \ \\
\includegraphics[clip,width=4.5cm]{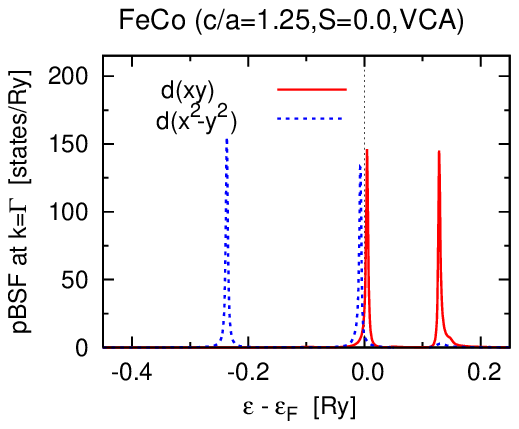} \\ \ \\
\includegraphics[clip,width=4.5cm]{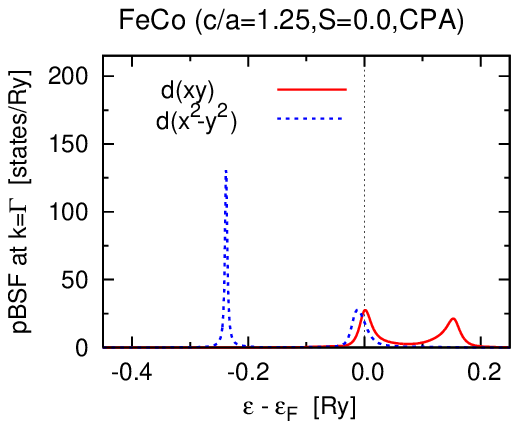}
\end{center}
\caption{Partial Bloch spectral function of $d_{x^2-y^2}$ and $d_{xy}$ states at the $\Gamma$ point in FeCo with $x=0.5$ and $c/a=1.25$: (a) results for the ordered structure $(S=1.0)$; (b) and (c) results for the disordered structure $(S=0.0)$ from the VCA and CPA, respectively.}
\label{fig3}
\end{figure}

To gain insight into the physical feature giving rise to the different behavior of the degree of order dependence of the MCA calculated using the VCA and CPA, we focus on the partial Bloch spectral functions of the $d_{x^2-y^2}$ and $d_{xy}$ orbital components in the minority-spin states at the $\Gamma$-point plotted in Fig. \ref{fig3}. According to Burkert {\it et al.}, these two orbital components have a major contribution to the MCA energy of tetragonal distorted FeCo alloys such that the magnetic easy axis coincides with the $c$-direction. This can be understood by expressing the energy variation arising within the second-order perturbation in terms of the spin-orbit interaction $\mathcal{H}_\mathrm{SO}$ \cite{bruno1989,wang1993},
\begin{equation}
\delta E = \sum^\mathrm{occ}_{n} \sum^\mathrm{unocc}_{n'} \frac{| \langle \bm{k}n_{\downarrow} | \mathcal{H}_\mathrm{SO} | \bm{k} n'_{\downarrow} \rangle |^2}{\varepsilon_{\bm{k}n\downarrow} - \varepsilon_{\bm{k}n'\downarrow}} \bigg|_{\bm{k}=\Gamma}
\label{eq1}
,\end{equation}
where $\varepsilon_{\bm{k} n \downarrow}$ denotes the eigenvalue of the nonperturbative eigenstate $| \bm{k}n_\downarrow \rangle$ in the minority-spin state. Thus, the projected components $\langle d_{xy} | \bm{k}n_\downarrow \rangle \cdot | d_{xy} \rangle$ and $\langle d_{x^2-y^2} | \bm{k}n_\downarrow \rangle \cdot | d_{x^2-y^2} \rangle$ mainly contribute to $\delta E$ and MCA energy. The restriction of the summation is applied to the occupied (unoccupied) state below (above) the Fermi level; therefore, eq. (\ref{eq1}) indicates that the MCA energy becomes large when the Fermi level lies between $d_{x^2-y^2}$ and $d_{xy}$ states. As shown in Fig. \ref{fig3}, the partial Bloch spectral function of FeCo $(x=0.5, c/a=1.25)$ for (a) ordered structure ($S=1.0$) and (b) disordered one ($S=0.0$) calculated using the VCA certainly supports this picture. The peaks of the spectra of the occupied $d_{x^2-y^2}$ and unoccupied $d_{xy}$ states are coupled to each other through the spin-orbit interaction, thereby inducing a large MCA. The MCA energy is more enhanced for $S=1.0$, since the energy separation between the occupied $d_{x^2-y^2}$ and unoccupied $d_{xy}$ states is smaller than that for $S=0.0$ calculated using the VCA. This is one of the reasons why the uniaxial anisotropy constants increase with $S$ in Fig. \ref{fig2}.

However, the spectra for $S=0.0$ obtained via the CPA shown in Fig. \ref{fig3}(c) also seem to meet this condition, although it should be noted here that in Fig. \ref{fig3}(c), the shapes of these spectra around the Fermi level are modified owing to electron scattering by the random arrangement of atoms. The width of the spectra increases compared with that obtained using the VCA as shown in Fig. \ref{fig3}(b). Some portions of the spectrum lie in the occupied state and others lie in the unoccupied states because of the wide energy distribution of each spectrum around the Fermi level. When we evaluate the energy variation $\delta E$ that results from mixing  $d_{xy}$ and $d_{x^2-y^2}$ states in the minority-spin state through the spin-orbit interaction by  using the nonperturbative Green's function obtained using the CPA and VCA based on the approach of Solovyev {\it et al.} \cite{solovyev1995}, we confirm that $\delta E$ calculated using the CPA is about one-half of that found using the VCA \cite{kota2012c}; therefore, this modification of the spectra contributes to decreasing the MCA energy in the bct FeCo alloy in Figs. \ref{fig1} and \ref{fig2}. Since in the CPA calculation, the energy widths of these two states increase with decreasing $S$, the $K_\mathrm{u}$ values from the CPA decrease much faster than those from the VCA when  $S$ decreases. Therefore, the giant MCA energy of the bct FeCo alloy can be realized under the quite strict condition concerning the $d_{xy}$ and $d_{x^2-y^2}$ states at the $\Gamma$-point.

In summary, we investigated the dependence of MCA energy on the degree of order of FeCo alloys by means of a first-principles calculation combined with the CPA. The results indicate that the MCA energy of FeCo alloys strongly depends on the degree of order under the optimal condition of the axial ratio $c/a=1.25$. This is because the modification of the spectral function resulting from  electron scattering in the chemical disordered structure has a considerable influence on the MCA energy under this condition, and so the $K_\mathrm{u}$ value of the FeCo alloy obtained using the CPA is about one-half of that obtained using the VCA; therefore, electron scattering has a significant influence on the MCA energy, which gives rise to the large difference between the VCA and CPA calculations in the small-$S$ region. Finally, we should stress that the high degree of order, in addition to the large distortion of the lattice, is a necessary condition for obtaining the large MCA from FeCo in experiments.

\acknowledgments
One of the authors (Y.K.) was supported by a Grant-in-Aid for JSPS Fellows (22-6092). This work was supported by JST under Collaborative Research Based on Industrial Demand ``High Performance Magnets: Towards Innovative Development of Next Generation Magnets''.

\clearpage



\clearpage

\begin{thebibliography}{99}
\bibitem{burkert2004}
T. Burkert, L. Nordstr\"om, O. Erikssonm, and O. Heinonen: Phys. Rev. Lett. {\bf 93} (2004) 027203.
\bibitem{klemmer1995}
T. Klemmer, D. Hoydick, H. Okumura, B. Zhang, and W. A. Soffa: Scripta Met. Mater. {\bf 33} (1995) 1792.
\bibitem{andersson2006}
G. Andersson, T. Burkert, P. Warnicke, M. Bj\"orck, B. Sanyal, C. Chacon, C. Zlotea, L. Nordst\"orm, P. Nordblad, and O. Eriksson: Phys. Rev. Lett. {\bf 96} (2006) 037205.
\bibitem{winkelmann2006}
A. Winkelmann, M. Przybylski, F. Luo, Y. Shi, and J. Barthel: Phys. Rev. Lett. {\bf 96} (2006) 257205.
\bibitem{warnicke2007}
P. Warnicke, G. Andersson, M. Bj\"orck, J. Ferre\'e, and P. Nordblad: J. Phys.: Condens. Matter {\bf 19} (2007) 226218.
\bibitem{luo2007}
F. Luo, X. L. Fu, A. Winkelmann, and M. Przbylski: Appl. Phys. Lett. {\bf 91} (2007) 262512.
\bibitem{yildiz2009}
F. Yildiz, M. Przybylski, X.-D. Ma, and J. Kirschner: Phys. Rev. B {\bf 80} (2009) 064415.
\bibitem{neise2011}
C. Neise, S. Sch\"onecker, M. Richterm K. Koepernik, and H. Eschrig: Phys. Status Solidi B {\bf 248} (2011) 2398.
\bibitem{hautier2012}
G. Hautier, A. Jain, and S. P. Ong: J. Mater. Sci. {\bf 47} (2012) 7317.
\bibitem{velicky1968}
B. Velick\'y, S. Kirkpatrick, and H. Ehrenreich: Phys. Rev. {\bf 175} (1968) 747.
\bibitem{elliott1974}
R. J. Elliott, J. A. Krumhansl, and P. L. Leath: Rev. Mod. Phys. {\bf 46} (1974) 465.
\bibitem{staunton2004}
J. B. Staunton, S. Ostanin, S. S. A. Razee, B. Gyorffy, L. Szunyogh, B. Ginatempo, and E. Bruno: J. Phys. C: Condens. Matter {\bf 16} (2004) S5623.
\bibitem{kota2012a}
Y. Kota and A. Sakuma: J. Appl. Phys. {\bf 111} (2012) 07A310.
\bibitem{kota2012b}
Y. Kota and A. Sakuma: J. Phys. Soc. Jpn. {\bf 81} (2012) 084705.
\bibitem{skriver1984}
H. L. Skriver: \emph{The LMTO Method} (Springer, Berlin, 1984).
\bibitem{turek1997}
I. Turek, V. Drchal, J. Kudrnovsk\'y, M. S\v{o}b, and P. Weinberger: \emph{Electronic Structure of Disordered Alloys, Surface and Interfaces} (Kluwer, Boston, 1997).
\bibitem{turek2008}
I. Turek, V. Drchal, and J. Kudrnovsk\'y: Philos. Mag. {\bf 88} (2008) 2787.
\bibitem{bozorth1951}
R. M. Bozorth: \emph{Ferromagnetism} (van Nostrand, New York, 1951).
\bibitem{bruno1989}
P. Bruno: Phys. Rev. B {\bf 39} (1989) 865.
\bibitem{wang1993}
D. Wang, R. Wu, and A. J. Freeman: Phys. Rev. B {\bf 47} (1993) 14932.
\bibitem{solovyev1995}
I. V. Solovyev, P. H. Dederichs, and I. Mertig: Phys. Rev. B {\bf 52} (1995) 13419.
\bibitem{kota2012c}
Y. Kota and A. Sakuma: in preparation for publication.
\end{thebibliography}
\end{document}